\documentclass{article}
\usepackage[table]{xcolor}
\usepackage[english]{babel}
\usepackage[utf8]{inputenc}
\usepackage{amsmath, amsfonts, graphicx, 
			color, hyperref, verbatim, natbib, 
            rotating, %endfloat, 
            authblk, tablefootnote}
            
\usepackage[english]{babel}
\usepackage[utf8]{inputenc}
\usepackage[colorinlistoftodos]{todonotes}
\usepackage{authblk}
\usepackage{booktabs}
\usepackage{pdflscape}

\usepackage[superscript,biblabel]{cite}
\usepackage[font=small,labelfont=bf]{caption}
\usepackage{bm}
\usepackage{soul}
\usepackage{empheq}
\usepackage{mdwlist}
\usepackage{lscape}
\usepackage{colortbl}
\usepackage{float}

\usepackage{titlesec}
\titlespacing*{\subsubsection}{0pt}{1\baselineskip}{0\baselineskip}

\title{Individualized Dynamic Prediction of Survival under Time-Varying Treatment Strategies}

\author[1, 2]{\textbf{Grigorios Papageorgiou}}
\author[2]{\bfseries Mostafa M. Mokhles}
\author[2]{\textbf{Johanna J. M. Takkenberg}} 
\author[1]{\textbf{Dimitris Rizopoulos}}

\affil[1]{\small{Department of Biostatistics, Erasmus Medical Center, Rotterdam, the Netherlands}}
\affil[2]{\small{Department of Thorax Surgery, Erasmus Medical Center, Rotterdam, the Netherlands}}
\date{}

\begin{document}
\maketitle

\begin{abstract}
Often in follow-up studies intermediate events occur in some patients, such as reinterventions or adverse events. These intermediate events directly affect the shapes of their longitudinal profiles. Our work is motivated by two studies in which such intermediate events have been recorded during follow-up. The first study concerns Congenital Heart Diseased (CHD) patients who were followed-up echocardiographically. Several patients were reintervened during follow-up. The second study concerns patients who participated in the SPRINT study and experienced adverse events during follow-up. In both studies we are interested in studying how the longitudinal evolutions change after the occurrence of the intermediate event and then utilize this information to improve the accuracy of the dynamic prediction for their risk.

To achieve this goal we propose here a flexible joint modeling framework for the longitudinal and time-to-event data that includes the intermediate event as a time-varying binary covariate in both the longitudinal and survival submodels. We consider a set of joint models that postulate different effects of the intermediate event in the longitudinal profile and the risk of the clinical endpoint, with different formulations for the association structure while allowing its functional form to change after the occurrence of the intermediate event. Based on these models we derive dynamic predictions of conditional survival probabilities which are adaptive to different scenarios with respect to the occurrence of the intermediate event. We evaluate the predictive accuracy of these predictions with a simulation study using the time-dependent area under the receiver operating characteristic curve and the expected prediction error adjusted to our setting. The results suggest that accounting for the changes in the longitudinal profiles and the instantaneous risk for the clinical endpoint is important, and improves the accuracy of the dynamic predictions.     
\end{abstract}

\section{Introduction}

Nowadays there is great interest in the medical field for procedures that facilitate precision medicine. In the context of follow-up studies, in which patients are monitored with several longitudinally measured parameters and biomarkers, physicians are interested in utilizing this information for predicting clinical endpoints. In this setting, joint models for longitudinal and survival outcomes provide a flexible framework to study the association between these outcomes and derive dynamic individualized predictions \citep{art1, art2, art3}.

The evaluation of the accuracy of these predictions obtained from joint models has gathered a lot of attention lately \citep{art4, art5, art6}. An important observation that has been made is that the accuracy of the derived predictions is influenced by the appropriate modeling of the subject-specific longitudinal profiles. In that regard, often in follow-up studies intermediate events occur in some patients that directly affect the shapes of their longitudinal evolutions. These may include events that are either directly in the control of the investigators, such as additional reinterventions, or may be not, such as adverse events that the patients may experience. While such intermediate events are common, very little work has been done in the direction of developing predictive tools that account for them and are adaptive to different scenarios with respect to their occurrence. To our knowledge only S\`ene et al. (2016) \citep{art7} investigated this topic in the context of prostate cancer recurrence and radiotherapy as an intermediate event. In their approach, however, they only considered the biomarker trajectories up to the occurrence of the intermediate event assuming extrapolation of the longitudinal profile thereafter. That is, changes in the shape of the longitudinal profile due to the occurrence of the intermediate event were not accounted for. Our goal is to show that utilizing the whole longitudinal trajectory, while capturing the changes to its shape due to intermediate events, can considerably improve the accuracy of such predictions. 

In our work we are motivated by two studies in which such intermediate events have been recorded during follow-up. The first study concerns 467 Congenital Heart Diseased (CHD) patients who underwent an initial pulmonary valve operation (i.e. Right Ventricular Outflow Reconstruction (RVOT), Pulmonary Valve Replacement) and were followed-up echocardiographically thereafter, at the Thorax Surgery department of Erasmus University Medical Center. Death is considered as the study endpoint while pulmonary gradient is the biomarker of interest which is believed to be related with the risk of death. During follow-up $65 \left(13.92\%\right)$ were reoperated and received a pulmonary allograft. In Figure 1 the pulmonary gradient evolutions of $4$ randomly selected patients, one from each combination between the reoperation and event status, are shown. Time of reoperation is depicted as a vertical red dashed line and it can be seen that when a patient is reoperated the pulmonary gradient drops. The interest in this study lies in the association between the pulmonary gradient and the risk of death but the main focus is to study the impact of reoperation on the risk of death both directly and indirectly (i.e. through its association with the pulmonary gradient) in order to develop predictive tools that can quantify the potential benefit of reoperation for future patients. The second study concerns 9361 persons who participated in the SPRINT study \citep{art8}. Subjects with increased cardiovascular risk, systolic blood pressure of 130 mm Hg or higher but without diabetes, were randomized to intensive or standard treatment. The composite primary outcome was myocardial infraction, acute coronary syndromes, stroke, heart failure or death from cardiovascular causes while systolic blood pressure is the biomarker of interest which was repeatedly measured. During follow-up $3424 \left(36.6\%\right)$ experienced serious adverse events. In Figure 2 the systolic blood pressure evolutions over time of the subjects who experienced serious adverse events and those who did not are depicted along with a loess curve (black thick line) for the average evolution over time for each group. The interest lies in assessing the impact of serious adverse events both in the systolic blood pressure evolution and the risk for the event of interest and exploiting it to derive individualized dynamic predictions for future patients with different scenarios with respect to the occurrence or not of serious adverse events. 

In both studies, physicians are interested in obtaining predictions of the respective primary clinical endpoints. However, to provide predictions of adequate accuracy, it will be required to carefully model the subject-specific longitudinal trajectories. Borrowing ideas from piecewise regression models, we achieve this by explicitly introducing the occurrence of these intermediate events as binary time-varying covariates in the specification of both the longitudinal and survival submodels of the joint model. The regression coefficient associated with this covariate can then capture changes, due to the occurrence of the intermediate event, in both the biomarker trajectory and the hazard for the event of interest. Furthermore we allow features of the biomarker trajectory, such as the rate of change to differ after the occurrence of the intermediate event. This allows us to estimate the impact of intermediate events as well as their specific features which then can be utilized in deriving dynamic predictions for a future patient under different scenarios: for example how the risk of a patient changes assuming different treatment strategies, such as no reintervention, reintervention now or reintervention at a later time-point.

The rest of the paper is structured as follows. Section 2 describes the formulation of the joint model in the presence of intermediate events. Section 3 presents the individualized dynamic predictions under different scenarios with respect to the occurrence of intermediate events and measures of predictive accuracy. In Section 4 we present the results from the analyses of the two motivating studies while in Section 5 we show the results of a simulation study. Finally, in Section 6 we close with a discussion.

\begin{minipage}{\linewidth}
\includegraphics{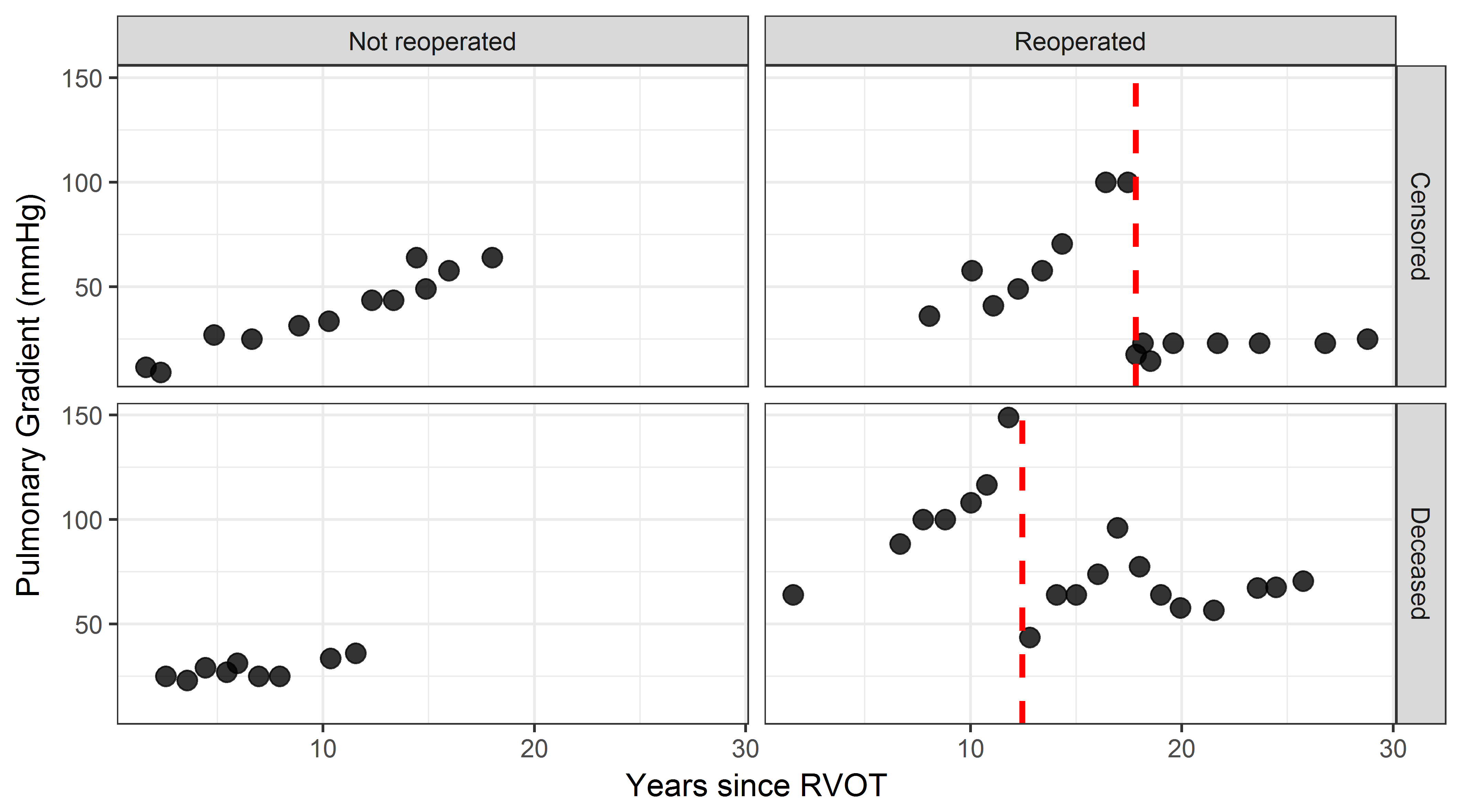}
\captionof{figure}{Pulmonary Gradient profile of 4 randomly selected patients, one from each of the following categories: Reoperated and Deceased, Reoperated and Censored, Not Reoperated and Deceased, Not Reoperated and Censored.}
\end{minipage}

\vspace{0.8cm}

\begin{minipage}{\linewidth}
\includegraphics{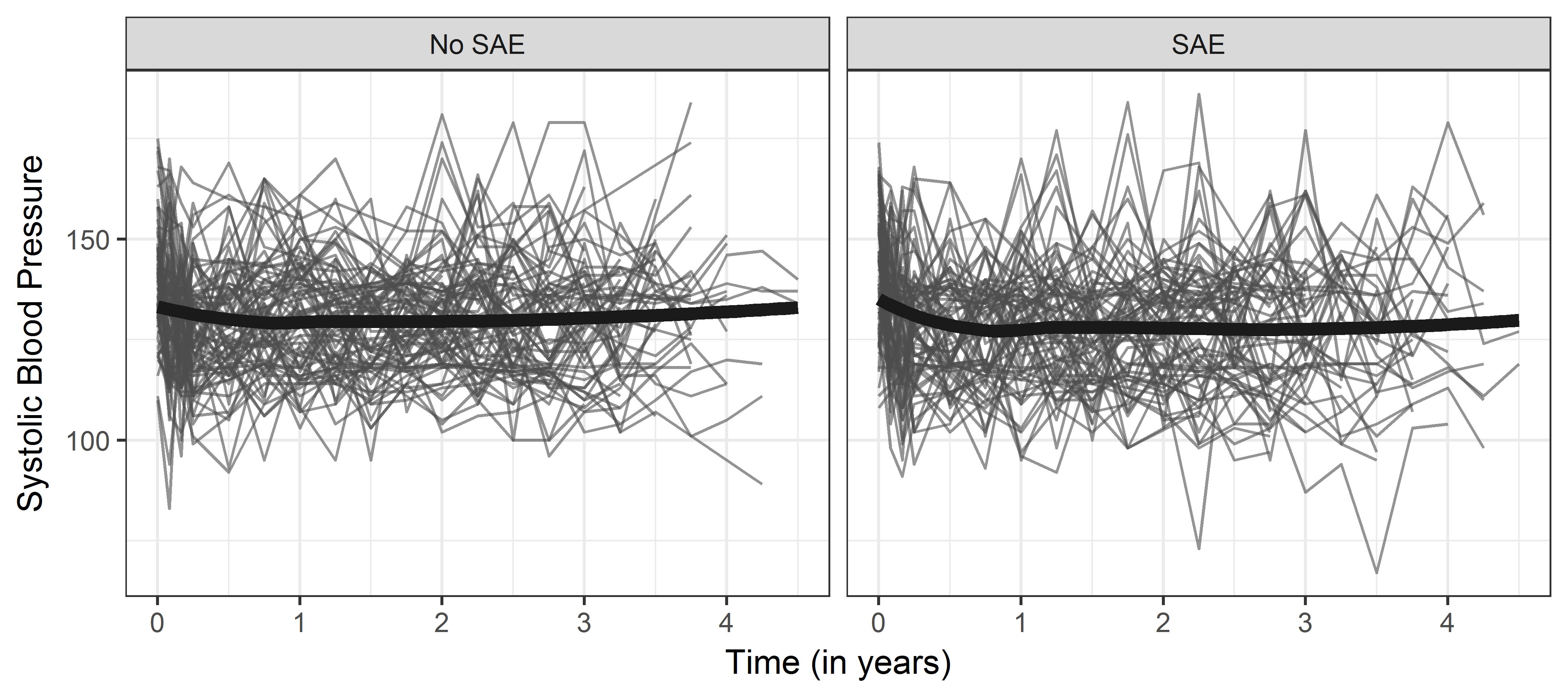}
\captionof{figure}{Individual evolutions and loess splines curves of systolic blood pressure for 160 randomly selected subjects with SAE and NO SAE.}
\end{minipage}

\section{Joint Model for Longitudinal and Time-to-Event Data with an Intermediate Event}

Assuming $n$ individuals under study, let $\mathcal{D}_{n} = \{T_{i}, \delta_{i}, \bm{y}_{i}, \rho_{i};\hspace{0.2cm} i = 1, \dots, n\}$ denote the sample from the target population, where $T_{i} = \min\left(T_{i}^{\star}, C_{i}\right)$ denotes the observed event time, which is defined as the minimum value between the true event time $T_{i}^{\star}$ and the censoring time $C_{i}$, and $\delta_{i} = I\left(T_{i}^{\star} \leq C_{i}\right)$ the event indicator with $I\left(\cdot\right)$ being the indicator function which equals $1$ if $T_{i}^{\star} \leq C_{i}$ and $0$ otherwise. Moreover, let $\rho_{i}$ denote the time to the intermediate event with a corresponding indicator $R_{i}\left(t\right) = I\left(t \geq \rho_{i}\right)$ at any time $t$ during follow-up, which takes the value $1$ if a subject experienced the intermediate event and $0$ otherwise. Furthermore, let $t_{i+} = \max\left(0, t_{ij} - \rho_{i}; j = 1, \dots, n_{i}\right)$ denote the time relative to the occurrence of the intermediate event and $\bm{y}_{i}$ be the vector of size $n_{i} \times 1$ of repeated measurements for the $i^{th}$ subject, with element $y_{ij}$ being the observed value of the longitudinal outcome at time point $t_{ij}, \hspace{0.2cm} j = 1, \dots, n_{i}$. We assume $\bm{y}_{i}$ to be a contaminated with measurement error version of the true and unobserved value of the longitudinal outcome at any time $t$: $\bm{y}_{i}\left(t\right) = \bm{\eta}_{i}\left(t\right) + \bm{\epsilon}_{i}\left(t\right)$ with $\bm{\eta}_{i}\left(t\right)$ denoting the true value of the longitudinal outcome at time $t$ and measurement error $\bm{\epsilon_{i}\left(t\right)} \sim \mathcal{N}\left(0, \sigma^{2}I_{n_{i}}\right)$. The true level of the longitudinal outcome is then formulated as: 

\begin{equation}\label{eq:1}
\eta_{i}\left(t\right) =
\left\{
\begin{array}{lr}
\bm{x}_{i}^{\top}\left(t\right)\bm{\beta} + \bm{z}_{i}^{\top}\left(t\right)\bm{b}_{i},& \hspace{0.4cm} 0 < t < \rho_{i} \vspace{0.4cm} \\
\bm{x}_{i}^{\top}\left(t\right)\bm{\beta} + \bm{z}_{i}^{\top}\left(t\right)\bm{b}_{i} + \tilde{\bm{x}}_{i}^{\top}\left(t\right)\tilde{\bm{\beta}} + \tilde{\bm{z}}_{i}^{\top}\left(t\right)\tilde{\bm{b}}_{i},& \hspace{0.4cm} t \geq \rho_{i},
\end{array}
\right.
\end{equation}
\linebreak
where $\bm{x}_{i}^{\top}\left(t\right)$ and $\bm{z}_{i}^{\top}\left(t\right)$ are design vectors for the fixed-effects regression coefficients $\bm{\beta}$ and the random-effects $\bm{b_{i}}$ respectively. Design vectors $\tilde{\bm{x}}_{i}^{\top}\left(t\right)$ and $\tilde{\bm{z}}_{i}^{\top}\left(t\right)$ include any function of the time-varying covariates $R_{i}\left(t\right)$ and $t_{i+}$, which describe the changes of the longitudinal trajectory after the occurrence of the intermediate event. These changes are then captured by the corresponing fixed-effects regression coefficients $\tilde{\bm{\beta}}$ allowing for subject-specific variation via the random-effects $\tilde{\bm{b}}_{i}$. The random-effects $\bm{b}_{i}$ and $\tilde{\bm{b}_{i}}$ are assumed to be normally distributed with mean zero and a $q \times q$ variance-covariance matrix $\bm{D}$.  

Depending on how the trajectory of the biomarker changes after the occurrence of the intermediate event, the specification of $\tilde{\bm{x}}_{i}^{\top}\left(t\right)$ and $\tilde{\bm{z}}_{i}^{\top}\left(t\right)$ may vary. Let $g\{R_{i}\left(t\right), t_{i+}\} = \tilde{\bm{x}}_{i}^{\top}\left(t\right)\tilde{\bm{\beta}} + \tilde{\bm{z}}_{i}^{\top}\left(t\right)\tilde{\bm{b}}_{i}$ denote the part of \eqref{eq:1} which describes the changes in the longitudinal profile after the occurrence of the intermediate event. Then in a setting as the one illustrated in Figure $1$, for the Pulmonary Gradient dataset, where the longitudinal trajectory is characterized by a seemingly linear evolution before and after the occurrence of the intermediate event, a steep drop at the occurrence of the intermediate event and a potential change in the slope after the occurrence of the intermediate event, function $g\{R_{i}\left(t\right), t_{i+}\}$ could be specified as $R_{i}\left(t\right)\left(\tilde{\beta}_{1} + \tilde{b}_{i1}\right) + t_{+}\left(\tilde{\beta}_{2} + \tilde{b}_{i2}\right)$. That is, the steep drop at the occurrence of the intermediate event will be captured by $\left(\tilde{\beta}_{1} + \tilde{b}_{i1}\right)$ and the change in the slope after the occurrence of the intermediate event will be captured by $\left(\tilde{\beta}_{2} + \tilde{b}_{i2}\right)$. On the other hand, in the setting of the SPRINT data where the longitudinal profiles show a nonlinear evolution over time without steep sudden changes, function $g\{R_{i}\left(t\right), t_{i+}\}$ could be specified as $\sum_{k = 1}^{K} \left(\tilde{\beta}_{k} + \tilde{b}_{ik}\right) B_{K}\left(t_{+}, \bm{k}\right)$ where $B_{K}\left(t_{+}, \bm{k}\right)$ denotes the $k^{\text{th}}$ basis function of a B-spline with knots $k_{1}, \dots, k_{Q}$. In that case the change in the slope of the nonlinear longitudinal trajectory after the occurrence of the intermediate event will be captured by $\left(\tilde{\beta}_{2} + \tilde{b}_{i2}\right)$ while there is no need to include $R_{i}\left(t\right)$. Generally, the functional form of $g\{R_{i}\left(t\right), t_{i+}\}$ may vary allowing for a broad range of mixed-effects models that can capture various types of changes in the longitudinal profile after the occurrence of the intermediate event.   

Let $\mathcal{H}_{i}\left(t, \rho_{i}\right) = \left[\eta_{i}\{s, \rho_{i}\left(s\right)\}, \hspace{0.2cm} 0 \leq s \leq t\right]$ denote the history of the longitudinal outcome up to time $t$. Note that in the definition of the history of the longitudinal outcome, we explicitly indicate that the true underlying value of the longitudinal outcome is also a function of the time to the intermediate event $\rho_{i}$. That is, to highlight that the subject-specific trajectory, $\eta_{i}\left(t\right)$, differs from the occurrence of the intermediate event onwards. Then the effects of the longitudinal outcome and the intermediate event, while adjusting for other covariates, on the risk for an event are quantified by utilizing a relative risk model of the form:

\begin{equation}\label{eq:2}
h_{i}\{t \mid \mathcal{H}_{i}\left(t, \rho_{i}\right), \bm{w}_{i}\} = 
\left\{
\begin{array}{lr}
h_{0}\left(t\right)\exp\left[\gamma^{\top} \bm{w}_{i} + f_{t < \rho_{i}}\{\mathcal{H}_{i}\left(t, \rho_{i}\right), \bm{b}_{i} \}^{\top}\bm{\alpha}\right],& \hspace{0.4cm} 0 < t < \rho_{i} \vspace{0.4cm} \\ 
h_{0}\left(t\right)\exp\left[\gamma^{\top} \bm{w}_{i} + R_{i}\left(t\right)\zeta + f_{t \geq \rho_{i}}\{\mathcal{H}_{i}\left(t, \rho_{i}\right), \bm{b}_{i} \}^{\top}\bm{\alpha}\right],& \hspace{0.4cm} t \geq \rho_{i},
\end{array}
\right.
\end{equation}
\linebreak
where $h_{0}\left(t\right)$ is the baseline risk function and $\bm{w}_{i}$ is a vector of baseline covariates with a corresponding vector of regression coefficients $\gamma$. The effect of the intermediate event on the risk is captured by the regression coefficient $\zeta$ which quantifies the change in risk from the the occurrence of the intermediate event onwards. Furthermore, the hazard of an event for patient $i$ at any time $t$ is associated with the subject-specific trajectory, $\eta_{i}\left(t\right)$, through $f\{(\mathcal{H}_{i}\left(t, \rho_{i}\right), \bm{b}_{i}\}$ which is a function of the history of the longitudinal outcome up to time $\mathcal{H}_{i}\left(t, \rho_{i}\right)$ and/or the vector of subject-specific effects $\bm{b}_{i}$. Function $f_{\left(t, \rho_{i}\right)}\{(\mathcal{H}_{i}\left(t, \rho_{i}\right), \bm{b}_{i}\}$ determines the association structure between the longitudinal and the time-to-event processes while the corresponding vector of regression coefficients $\bm{\alpha}$ quantifies the magnitude of the association. Several functional forms for the specification of the association structure have been used in the literature, such as the current value, current slope and the cumulative effect \citep{art11}. The functional form of the association structure is an important feature of the joint model formulation, especially with regard to the accuracy of the dynamic predictions \citep{art4, art9, art10}. Hence, to allow for more flexibility we explicitly allow for the functional form of the association structure to differ before and after the occurrence of the intermediate event. In general, any functional form can be used for $f_{t \geq \rho_{i}}\{\mathcal{H}_{i}\left(t, \rho_{i}\right), \bm{b}_{i} \}$ and $f_{t < \rho_{i}}\{\mathcal{H}_{i}\left(t, \rho_{i}\right), \bm{b}_{i} \}$ including, of course, the case where the association structure remains the same and $f_{t \geq \rho_{i}}\{\mathcal{H}_{i}\left(t, \rho_{i}\right), \bm{b}_{i} \} = f_{t < \rho_{i}}\{\mathcal{H}_{i}\left(t, \rho_{i}\right), \bm{b}_{i} \}$.   

\section{Individualized Dynamic Predictions with Time-varying Intermediate Events}

\subsection{Dynamic Predictions}

Based on the joint model fitted in the sample $\mathcal{D}_{n} = \{T_{i}, \delta_{i}, \bm{y}_{i}, \rho_{i};\hspace{0.2cm} i = 1, \dots, n\}$ from the target population, dynamic predictions for a new subject $j$ from the same population can be derived up to a future time of interest $u > t$ given his/hers biomarker history $\mathcal{H}_{j}\left(t\right) = \left[\eta_{j}\{s, R_{j}\left(s\right)\}, \hspace{0.2cm} 0 \leq s \leq t\right]$. Let $\mathcal{Y}_{j}\left(t\right) = \{y_{j}\left(t_{jl}\right); 0 \leq t_{jl} \leq t, l = 1, \dots, n_{j}\}$ denote the history of observed biomarker values taken up to time $t$ for patient $j$, then under the Bayesian joint model framework these predictions can be estimated using the corresponding posterior predictive distributions, namely:

\begin{equation}\label{eq:3}
\begin{array}{rcl}
\pi_{j}\left(u\mid t\right) &=& Pr\{T_{j}^{\star}\geq u \mid T_{j}^{\star} > t, \mathcal{Y}_{j}\left(t\right), \bm{\theta}\} \vspace{0.4cm} \\  
&=& \displaystyle\int \frac{S_{j}\{u \mid \mathcal{H}_{j}\left(u, \bm{b}_{j}\right), \bm{\theta}\}} {S_{j}\{t \mid \mathcal{H}_{j}\left(t, \bm{b}_{j}\right), \bm{\theta}\}} p\{\bm{b}_{j}\mid T_{j}^{\star} > t, \mathcal{Y}_{j}\left(t\right), \bm{\theta}\}d\bm{b}_{j} \vspace{0.4cm} \\ 
&=& \displaystyle\int \exp\left[-\displaystyle\int_{t}^{u}h_{j}\{s \mid \mathcal{H}_{j}\left(u, \bm{b}_{j}\right)\}ds\right] p\{\bm{b}_{j}\mid T_{j}^{\star} > t, \mathcal{Y}_{j}\left(t\right), \bm{\theta}\}d\bm{b}_{j}, \hspace{0.2cm} t \leq s \leq u.
\end{array}
\end{equation}
\linebreak
Expressing the fraction term in \eqref{eq:3} as $\exp\{-\int_{t}^{u}h_{i}\left(s\right)ds\}, t \leq s \leq u$ has two main advantages. First, it reduces the computational time required, since the denominator part $S_{j}\{t \mid \mathcal{H}_{j}\left(t, \bm{b}_{j}\right), \bm{\theta}\}$ does not need to be computed anymore. Second, it improves the precision of numerical integration. The latter benefit is due to the fact that by re-expressing the fraction term in \eqref{eq:3} as such, an adaptive Gauss-Kronrod scheme can be deployed for the numerical computation of smaller regions of the target interval. This improves the precision of Gaussian quadrature, since the quadrature points are spent for smaller regions of the interval.

Incorporating the time-varying covariate $R_{i}\left(t\right)$ in both the longitudinal and relative risk submodels of the joint model allows us to evaluate how the occurrence of the intermediate event of interest at a future time point will influence the individualized risk predictions, for subjects who have not experienced the intermediate event by time $t$. The main difference of our approach when compared with the approach of S\`ene et al. (2016) \citep{art7}, is that we assume that both the instantaneous risk of the primary endpoint and the longitudinal profile change after the occurrence of the intermediate event, whereas they assumed an extrapolated longitudinal profile, instead. More specifically, by assuming an extrapolated longitudinal profile, S\`ene et al. (2016), were more interested in assessing how predictions change with and without a second treatment whereas we are more interested in studying how individualized risk predictions are influenced by intermediate events, such as reintervention or adverse events, by explicitly allowing for changes in both the longitudinal and survival submodels. 

That is, different scenarios regarding the time of the intermediate event may lead to changes in the risk captured via different individual dynamic predictions accordingly. More specifically, for a future time of interest $u > t$ different assumptions can be made: $1\left.\right)$ The patient experiences an intermediate event immediately $\rho_{i} = t$ or at a time point within the time-interval of prediction $t \leq \rho_{i} \leq u$. $2\left.\right)$ The patient does not experience an intermediate event within the time-interval of prediction $\rho_{i} > u$. The individualized dynamic predictions in \eqref{eq:4} are then further dependent on the scenario of choice:

\begin{equation}\label{eq:4}
\pi_{j}\left(u\mid t, \rho_{j}\right) = \left\{
\begin{array}{rr}
\displaystyle\int \exp\{-\displaystyle\int_{t}^{u}h_{j}\left(s \mid \rho_{j} \leq u\right)ds\} \prod_{l}\{p\left(y_{j} \mid \rho_{j} \leq u, b_{j}\right)\} p\left(\bm{b}_{j}\mid \bm{\theta}\right)d\bm{b}_{j} &, \hspace{0.2cm} t \leq s \leq u \vspace{0.4cm} \\
\displaystyle\int \exp\{-\displaystyle\int_{t}^{u}h_{j}\left(s \mid \rho_{j} > u\right)ds\} \prod_{l}\{p\left(y_{j} \mid \rho_{j} > u, b_{j}\right)\} p\left(\bm{b}_{j}\mid \bm{\theta}\right)d\bm{b}_{j} &, \hspace{0.2cm} t \leq s \leq u. 
\end{array}
\right.
\end{equation}

In \eqref{eq:4} $\prod_{l}\{p\left(y_{j} \mid \rho_{j} \leq u, b_{j}\right)\}$ and $\prod_{l}\{p\left(y_{j} \mid \rho_{j} > u, b_{j}\right)\}$ are multivariate Gaussian joint densities for the longitudinal responses with means $\bm{x}_{j}^{\top}\left(t\right)\bm{\beta} + R_{j}\left(t\right)\beta_{R} + t_{+}\beta_{t_{+}} + \bm{z}_{j}^{\top}\left(t\right)\bm{b_{j}} + R_{j}\left(t\right)b_{jR} + t_{+}b_{jt_{+}}$ and $\bm{x}_{j}^{\top}\left(t\right)\bm{\beta} + \bm{z}_{j}^{\top}\left(t\right)\bm{b_{j}}$ respectively and variance-covariance matrices $\sigma^{2}I_{n_{j}}$. $p\left(\bm{b}_{j}\mid \bm{\theta}\right)$ is a multivariate Gaussian density function with mean $0$ and variance covariance matrix $D$.

To estimate $\pi_{j}\left(u\mid t, \rho_{j}\right)$ a Monte Carlo scheme is employed where a large set of $\theta^{m}\left(m = 1, \dots, M \right)$ and $b_{j}^{m}\left(m = 1, \dots, M \right)$ are sampled from their posterior distributions and then used to compute $\pi_{j}^{m}\left(u\mid t, \rho_{j}\right)$. The median value of $\pi_{j}^{m}\left(u\mid t, \rho_{j}\right)$ is the point estimate and the $2.5\%$ and $97.5\%$ percentiles give a $95\%$ credible interval.  

\subsection{Evaluation of predictive performance}

To assess the performance of the individualized dynamic predictions described in the previous section, we will work under a similar framework as the one presented in \citep{art4}. More specifically, we will use the time-dependent area under the receiver operating characteristic curve (AUC) and the expected prediction error (PE), adapted for the presence of intermediate events.

Under the framework presented in Sections 2 and 3.1 a rule can be defined using the individualized dynamic predictions $\pi_{j}\left(u = t + \Delta t\mid t \right)$ while utilizing the available longitudinal measurements up to $t$, $\mathcal{Y}_{j}\left(t\right)$. More specifically, a subject $j$ can be termed as either to experience the event $\pi_{j}\left(u = t + \Delta t\mid t \right) \leq c$ or not $\pi_{j}\left(u = t + \Delta t\mid t \right) > c$ within a clinically relevant time interval $\left(t, \Delta t\right]$, with $c \in \left[0, 1\right]$. That is, for a pair of subjects which is randomly chosen $\{i, j\}$ for both of which the measurements up to $t$ are provided, the AUC which is calculated for varying values of $c$ is a measure of the discriminative capability of the assumed model and is given by:

\begin{equation}\label{eq:5}
\text{AUC}\left(t, \Delta t\right) = \text{Pr}\left[\pi_{i}\left(t + \Delta t \mid t\right) < \pi_{j}\left(t + \Delta t \mid t \right) \mid \{T_{i}^{\star} \in \left(t, t + \Delta t\right]\} \cap \{T_{j}^{\star} > t + \Delta t\}\right],
\end{equation}
\linebreak
which intuitively means that we expect the assumed model to give higher probability of surviving longer than the time interval of interest $\left(t + \Delta t\right]$ to the subject who did not experience the event (in this case subject $j$).

However, in the presence of intermediate events the dynamic predictions change depending on whether a subject experienced the intermediate event or not. That is, the AUC in \eqref{eq:5} changes to:

\begin{equation}\label{eq:6}
\resizebox{.9\hsize}{!}{$
\text{AUC}\left(t, \Delta t\right) = \left\{
\begin{array}{rr} \text{Pr}\left[\pi_{i}\left(t + \Delta t \mid t, \rho_{i} > t\right) < \pi_{j}\left(t + \Delta t \mid t, \rho_{i} > t\right) \mid \{T_{i}^{\star} \in \left(t, t + \Delta t\right]\} \cap \{T_{j}^{\star} > t + \Delta t\}\right], &  0 < t < \rho_{i} \vspace{0.4cm}\\ 
\text{Pr}\left[\pi_{i}\left(t + \Delta t \mid t, \rho_{i} \leq t\right) < \pi_{j}\left(t + \Delta t \mid t, \rho_{i} \leq t\right) \mid \{T_{i}^{\star} \in \left(t, t + \Delta t\right]\} \cap \{T_{j}^{\star} > t + \Delta t\}\right], &  t \geq \rho_{i}.  
\end{array}
\right.
$}
\end{equation}

Estimation of $\text{AUC}\left(t, \Delta t\right)$ is based in counting the concordant pairs of subjects by appropriately distinguishing between the comparable and the non-comparable (due to censoring) pairs of subjects at time $t$. More specifically, the following decomposition is used:

\begin{equation}\nonumber
\hat{\text{AUC}}\left(t, \Delta t\right) = \hat{\text{AUC}}_{1}\left(t, \Delta t\right) + \hat{\text{AUC}}_{2}\left(t, \Delta t\right) + \hat{\text{AUC}}_{3}\left(t, \Delta t\right) + \hat{\text{AUC}}_{4}\left(t, \Delta t\right).
\end{equation}
\linebreak
Term $\hat{\text{AUC}}_{1}\left(t, \Delta t\right)$ refers to the pairs of subjects who are comparable, 

\begin{equation}\nonumber
\Omega_{ij}^{\left(1\right)}\left(t\right) = 
\begin{cases}
\left[\{T_{i} \in \left(t, t + \Delta t\right]\} \cap \{\delta_{i} = 1\} \cap \{0 < t < \rho_{i}\}\right] \cap \left[\{T_{j} > t + \Delta t\} \cap \{0 < t < \rho_{j} \}\right] \vspace{0.4cm} \\
\left[\{T_{i} \in \left(t, t + \Delta t\right]\} \cap \{\delta_{i} = 1\} \cap \{t \geq \rho_{i} \}\right] \cap \left[\{T_{j} > t + \Delta t\} \cap \{t \geq \rho_{j} \}\right],
\end{cases}
\end{equation}
\linebreak
where $i, j = 1, \dots, n$ with $i \neq j$. We can then estimate and compare the survival probabilities $\pi_{i}\left(t + \Delta t \mid t, t \geq \rho_{i}\right)$ and $\pi_{i}\left(t + \Delta t \mid t, t \geq \rho_{i}\right)$ for subjects $i$ and $j$ who did not experienced the intermediate event and $\pi_{i}\left(t + \Delta t \mid t, 0 < t < \rho_{i}\right)$ and $\pi_{i}\left(t + \Delta t \mid t, 0 < t < \rho_{i}\right)$ for subjects who experienced the intermediate event. Then $\hat{\text{AUC}}_{1}\left(t, \Delta t\right)$ is the proportion of concordant subjects out of the set of comparable subjects at time $t$:

\begin{equation}\nonumber
\begin{aligned}
\hat{\text{AUC}}_{1}\left(t, \Delta t\right) &= \frac{\sum\limits_{i \in \mathcal{A}} \sum\limits_{j \neq i \in \mathcal{A}} I\{\hat{\pi}_{i}\left(t + \Delta t \mid t, t < \rho_{i}\right) < \hat{\pi}_{j}\left(t + \Delta t \mid t, t < \rho_{j}\right)\} \times I\{\Omega_{ij}^{\left(1\right)}\left(t\right)\}}{\sum\limits_{i \in \mathcal{A}} \sum\limits_{j \neq i \in \mathcal{A}} I\{\Omega_{ij}^{\left(1\right)}\left(t\right)\}} \vspace{0.4cm} \\
&+ \frac{\sum\limits_{i \in \mathcal{B}} \sum\limits_{j \neq i \in \mathcal{B}} I\{\hat{\pi}_{i}\left(t + \Delta t \mid t, t \geq \rho_{i}\right) < \hat{\pi}_{j}\left(t + \Delta t \mid t, t \geq \rho_{j}\right)\} \times I\{\Omega_{ij}^{\left(1\right)}\left(t\right)\}}{\sum\limits_{i \in \mathcal{B}} \sum\limits_{j \neq i \in \mathcal{B}} I\{\Omega_{ij}^{\left(1\right)}\left(t\right)\}},
\end{aligned}
\end{equation}
\linebreak
where $\mathcal{A} = \{i, j: t < \rho_{i}; i, j = 1, \dots, n\}$ and $\mathcal{B} = \{i, j: t \geq \rho_{i}; i, j = 1, \dots, n\}$.

The remaining terms, $\hat{\text{AUC}}_{2}\left(t, \Delta t\right)$, $\hat{\text{AUC}}_{3}\left(t, \Delta t\right)$ and $\hat{\text{AUC}}_{4}\left(t, \Delta t\right)$ refer to the pairs of subjects who due to censoring cannot be compared, namely

\begin{equation}\nonumber
\begin{array}{l}
\Omega_{ij}^{\left(2\right)}\left(t\right) = 
\begin{cases}
\left[\{T_{i} \in \left(t, t + \Delta t\right]\} \cap \{\delta_{i} = 0\} \cap \{0 < t < \rho_{i}\}\right] \cap \left[\{T_{j} > t + \Delta t\} \cap \{0 < t < \rho_{j} \}\right] \vspace{0.4cm} \\
\left[\{T_{i} \in \left(t, t + \Delta t\right]\} \cap \{\delta_{i} = 0\} \cap \{t \geq \rho_{i} \}\right] \cap \left[\{T_{j} > t + \Delta t\} \cap \{t \geq \rho_{j} \}\right],
\end{cases} \vspace{0.8cm} \\

\Omega_{ij}^{\left(3\right)}\left(t\right) = 
\begin{cases}
\left[\{T_{i} \in \left(t, t + \Delta t\right]\} \cap \{\delta_{i} = 1\} \cap \{0 < t < \rho_{i}\}\right] \cap \left[\{T_{i} < T_{j} \leq t + \Delta t\} \cap \{0 < t < \rho_{j}\} \cap \{\delta_{j} = 0\}\right] \vspace{0.4cm} \\
\left[\{T_{i} \in \left(t, t + \Delta t\right]\} \cap \{\delta_{i} = 1\} \cap \{t \geq \rho_{i} \}\right] \cap \left[\{T_{i} < T_{j} \leq t + \Delta t\} \cap \{t \geq \rho_{j} \} \cap \{\delta_{j} = 0\}\right],
\end{cases} \vspace{0.8cm} \\

\Omega_{ij}^{\left(4\right)}\left(t\right) = 
\begin{cases}
\left[\{T_{i} \in \left(t, t + \Delta t\right]\} \cap \{\delta_{i} = 0\} \cap \{0 < t < \rho_{i}\}\right] \cap \left[\{T_{i} < T_{j} \leq t + \Delta t\} \cap \{0 < t < \rho_{j}\} \cap \{\delta_{j} = 0\}\right] \vspace{0.4cm} \\
\left[\{T_{i} \in \left(t, t + \Delta t\right]\} \cap \{\delta_{i} = 0\} \cap \{t \geq \rho_{i} \}\right] \cap \left[\{T_{i} < T_{j} \leq t + \Delta t\} \cap \{t \geq \rho_{j} \} \cap \{\delta_{j} = 0\}\right],
\end{cases}
\end{array}
\end{equation}
\linebreak
which contribute to the overall AUC after being appropriately weighted with the probability that they would be comparable:

\begin{equation}\nonumber
\begin{aligned}
\hat{\text{AUC}}_{m}\left(t, \Delta t\right) &= \frac{\sum\limits_{i \in \mathcal{A}} \sum\limits_{j \neq i \in \mathcal{A}} I\{\hat{\pi}_{i}\left(t + \Delta t \mid t, t < \rho_{i}\right) < \hat{\pi}_{j}\left(t + \Delta t \mid t, t < \rho_{j}\right)\} \times I\{\Omega_{ij}^{\left(m\right)}\left(t\right)\} \times \hat{\nu}_{ij}^{\left(m\right)}}{\sum\limits_{i \in \mathcal{A}} \sum\limits_{j \neq i \in \mathcal{A}} I\{\Omega_{ij}^{\left(1\right)}\left(t\right)\} \times \hat{\nu_{ij}^{\left(m\right)}}} \vspace{0.4cm} \\
&+ \frac{\sum\limits_{i \in \mathcal{B}} \sum\limits_{j \neq i \in \mathcal{B}} I\{\hat{\pi}_{i}\left(t + \Delta t \mid t, t \geq \rho_{i}\right) < \hat{\pi}_{j}\left(t + \Delta t \mid t, t \geq \rho_{j}\right)\} \times I\{\Omega_{ij}^{\left(1\right)}\left(t\right)\} \times \hat{\nu}_{ij}^{\left(m\right)}}{\sum\limits_{i \in \mathcal{B}} \sum\limits_{j \neq i \in \mathcal{B}} I\{\Omega_{ij}^{\left(1\right)}\left(t\right)\} \times \hat{\nu}_{ij}^{\left(m\right)}},
\end{aligned}
\end{equation}
\linebreak
with $m = 2, 3, 4$ and 

\begin{equation}\nonumber
\begin{array}{l}
\hat{\nu}_{ij}^{\left(2\right)} = \begin{cases}
\begin{array}{lr}
1 - \hat{\pi}_{i}\left(t + \Delta t \mid T_{i}, t < \rho_{i}\right), & i \in \mathcal{A} \vspace{0.4cm} \\
1 - \hat{\pi}_{i}\left(t + \Delta t \mid T_{i}, t \geq \rho_{i}\right), & i \in \mathcal{B}, 
\end{array}
\end{cases}
 \vspace{0.8cm} \\

\hat{\nu}_{ij}^{\left(3\right)} = \begin{cases}
\begin{array}{lr}
\hat{\pi}_{j}\left(t + \Delta t \mid T_{j}, t < \rho_{j}\right), & j \in \mathcal{A} \vspace{0.4cm} \\
\hat{\pi}_{j}\left(t + \Delta t \mid T_{j}, t \geq \rho_{j}\right), & j \in \mathcal{B}, 
\end{array}
\end{cases}
 \vspace{0.8cm} \\

\hat{\nu}_{ij}^{\left(4\right)} = \begin{cases}
\begin{array}{lr}
\{1 - \hat{\pi}_{i}\left(t + \Delta t \mid T_{i}, t < \rho_{i}\right)\} \times \hat{\pi}_{j}\left(t + \Delta t \mid T_{j}, t < \rho_{j}\right), & i,j \in \mathcal{A} \vspace{0.4cm} \\
\{1 - \hat{\pi}_{i}\left(t + \Delta t \mid T_{i}, t \geq \rho_{i}\right)\} \times \hat{\pi}_{j}\left(t + \Delta t \mid T_{j}, t \geq \rho_{j}\right), & i,j \in \mathcal{B}. 
\end{array}
\end{cases}
\end{array}
\end{equation}
\linebreak

The expected error of predicting future events can be used to assess the accuracy of individualized dynamic predictions. Similarly as for the AUC, to account for the dynamic nature of the longitudinal outcome, we focus our interest in predicting events that occur at a time point $u > t$ given the information available up to time $t$, $\mathcal{Y}_{j}\left(t\right)$. Let $N_{j}\left(t\right) = I\left(T_{i}^{\star} > t\right)$ denote the event status of subject $j$ at time $t$. Using the square loss function the expected prediction error then is:

\begin{equation}\label{eq:7}
\text{PE}\left(u \mid t\right) = E\left[\{N_{j}\left(u\right) - \pi_{j}\left(u \mid t\right)\}^{2}\right],
\end{equation}
\linebreak
where the expectation is taken with respect to the distribution of the event times. Adapting the above to the framework of intermediate events, \eqref{eq:7} can be re-expressed as:

\begin{equation}\label{eq:8}
\text{PE}\left(u \mid t\right) = \left\{
\begin{array}{lr}
E\left[\{N_{j}\left(u\right) - \pi_{j}\left(u \mid t, \rho_{i} > t\right)\}^{2}\right], & \hspace{0.2cm} 0 < t < \rho_{i} \vspace{0.4cm} \\ 
E\left[\{N_{j}\left(u\right) - \pi_{j}\left(u \mid t, \rho_{i} \leq t\right)\}^{2}\right], & \hspace{0.2cm} t \geq \rho_{i},
\end{array}
\right.
\end{equation}
\linebreak
where for each case the corresponding individualized dynamic predictions showed in \eqref{eq:4} are used. The estimate of $\text{PE}\left(u \mid t\right)$ as proposed by Henderson et al. (2002) \citep{art12} and adjusted for the presence of intermediate events is given by:

\begin{equation}\nonumber
\begin{aligned}
\hat{\text{PE}}\left(u \mid t, \rho_{i}\right) &= \{n\left(t, t < \rho_{i}\right)\}^{-1}\sum\limits_{i \in \mathcal{A}, T_{i} \geq t} I\left(T_{i} \geq u\right) \{1 - \hat{\pi}_{i}\left(u \mid t, t < \rho_{i}\right)\}^{2} + \delta_{i}I\left(T_{i} < u\right) \{0 - \hat{\pi}_{i}\left(u \mid t, t < \rho_{i}\right)\}^{2} \vspace{0.4cm} \\
&+ \left(1 - \delta_{i}\right) I\left(T_{i} < u\right) \left[\hat{\pi}_{i}\left(u \mid T_{i}, t < \rho_{i}\right) \{1 - \hat{\pi}_{i}\left(u \mid t, t < \rho_{i}\right)\}^{2} \{1 - \hat{\pi}_{i}\left(u \mid T_{i}, t < \rho_{i}\right)\}\right. \vspace{0.4cm} \\
&\times \left.\{0 - \hat{\pi}_{i}\left(u \mid t, t < \rho_{i}\right)\}^{2}\right] + \{n\left(t, t \geq \rho_{i}\right)\}^{-1}\sum\limits_{i \in \mathcal{B}, T_{i} \geq t} I\left(T_{i} \geq u\right) \{1 - \hat{\pi}_{i}\left(u \mid t, t \geq \rho_{i}\right)\}^{2} \vspace{0.4cm} \\ 
&+ \delta_{i}I\left(T_{i} < u\right) \{0 - \hat{\pi}_{i}\left(u \mid t, t \geq \rho_{i}\right)\}^{2} \left(1 - \delta_{i}\right) I\left(T_{i} < u\right) \left[\hat{\pi}_{i}\left(u \mid T_{i}, t \geq \rho_{i}\right) \{1 - \hat{\pi}_{i}\left(u \mid t, t \geq \rho_{i}\right)\}^{2}\right. \vspace{0.4cm} \\ 
&\times \left.\{1 - \hat{\pi}_{i}\left(u \mid T_{i}, t \geq \rho_{i}\right)\}\{0 - \hat{\pi}_{i}\left(u \mid t, t < \rho_{i}\right)\}^{2}\right], 
\end{aligned}
\end{equation}
\linebreak
where $n\left(t, t < \rho_{i}\right)$ and $n\left(t, t \geq \rho_{i}\right)$ denote the number of subjects still at risk at time $t$, who have not/have experienced the intermediate event respectively.

\section{Analysis of Pulmonary Gradient and SPRINT trial data}

\subsection{Pulmonary Gradient dataset}

The Pulmonary Gradient dataset was introduced in Section 1. Our goal, is to investigate the association between the pulmonary gradient and the risk of death, how reoperation as an intermediate event changes the evolution of the pulmonary gradient and the instantaneous risk for death, and then to utilize this information to derive individualized dynamic predictions under different scenarios with respect to a future time of reoperation. 

In Figure 1, the evolutions of pulmonary gradient for reoperated and non reoperated patients are depicted, where it is shown that for the case of reoperated patients the profiles show a linear increasing trend which drops at the moment of reoperation and then continues to increase whereas for the case of non reoperated patients the profiles show a linear increasing trend. Therefore, for this outcome, we assumed a linear mixed-effects submodel including a linear effect of time, a drop at the moment of reoperation and a change in slope after the occurrence of reoperation in both the fixed-effects and random-effects parts of the model while correcting for baseline differences in age and gender. A preliminary analysis suggested that assuming nonlinear effects of time did not improve the fit of the model to the data. Hence, we used the following specification for the mixed-effects model: 

\begin{equation}\label{eq:9}
\resizebox{.95\hsize}{!}{$
PG_{i}\left(t\right) = \left\{
\begin{array}{lr}
\left(\beta_{0} + b_{i0}\right) + \left(\beta_{1} + b_{i1}\right)\times t + \beta_{4}Age + \beta_{5}Gender + \epsilon_{i}\left(t\right), & \hspace{0.2cm} 0 < t < \rho_{i} \vspace{0.4cm} \\
\left(\beta_{0} + b_{i0}\right) + \left(\beta_{1} + b_{i1}\right)\times t + \left(\tilde{\beta}_{2} + \tilde{b}_{i2}\right)\times R_{i}\left(t\right) + \left(\tilde{\beta}_{3} + \tilde{b}_{i3}\right) \times t_{+} + \beta_{4}Age + \beta_{5}Gender + \epsilon_{i}\left(t\right), & \hspace{0.2cm} t \geq \rho_{i}.
\end{array}
\right.
$}
\end{equation}
\linebreak
where $PG_{i}\left(t\right)$ are the measurements of pulmonary gradient, $R_{i}\left(t\right)$ is a binary time-dependent indicator of reoperation and $t_{+} = \max\left(0, t_{ij} - \rho_{ij}; j = 1, \dots, n_{i}\right)$ is the time relative to reoperation.

To investigate the association between the pulmonary gradient and the risk of death, we postulated relative risk submodels with different parametrizations for the pulmonary gradient. The baseline hazard was expressed as a B-splines function. We also corrected for age and gender and assumed reoperation to have a direct effect on the hazard. Based on a preliminary analysis we assessed various formulations for the association structure. However, only functional forms including the slope of the pulmonary gradient were found to be stronger. Assuming a different functional form after the occurrence of reoperation did not improve the fit of the model to the data. Therefore, we present the joint models that include the slope of the pulmonary gradient along with the joint model that assumes an association with the current value of the pulmonary gradient for the shake of comparison, since it is the most common form of the association structure:   

\begin{equation}\nonumber
\begin{array}{rcl}
\vspace{0.4cm}
\text{M}1: h_{i}\left(t\right) &=& h_{0}\left(t\right)\exp\{\gamma_{1}\text{Age}_{i} + \gamma_{2}\text{Gender}_{i} + \zeta R_{i}\left(t\right) + \alpha_{1}\eta_{PGi}\left(t\right)\}, \\ 
\vspace{0.4cm}
\text{M}2: h_{i}\left(t\right) &=& h_{0}\left(t\right)\exp\{\gamma_{1}\text{Age}_{i} + \gamma_{2}\text{Gender}_{i} + \zeta R_{i}\left(t\right) + \alpha_{1}\eta_{PGi}\left(t\right) + \alpha_{2}\frac{d}{dt}\eta_{PGi}\left(t\right)\}, \\ \vspace{0.4cm}
\text{M}3: h_{i}\left(t\right) &=& h_{0}\left(t\right)\exp\{\gamma_{1}\text{Age}_{i} + \gamma_{2}\text{Gender}_{i} + \zeta R_{i}\left(t\right) + \alpha_{2}\frac{d}{dt}\eta_{PGi}\left(t\right)\}. \\
\end{array}
\end{equation}

Table 1 summarizes the parameter estimates and the 95\% credibility intervals of the longitudinal submodel that was used for the Pulmonary Gradient dataset. Table 2 summarizes the parameter estimates and the 95\% credibility intervals of the survival submodels based on the three joint models fitted to the Pulmonary Gradient dataset. As shown in Table 2, the association of the pulmonary gradient with the instantaneous risk of death was weak regardless the functional form for the association structure. The strongest association in magnitude was found when using the slope parametrization both before and after the occurrence of the intermediate event.   

\rowcolors{2}{gray!6}{white}
\begin{table}[!h]

\caption{\label{tab:}Estimated coefficients and $95\%$ credibility intervals for the parameters of the longitudinal submodel fitted to the Pulmonary Gradient dataset.}
\centering
\begin{tabular}[t]{lrl}
\hiderowcolors
\toprule
  & Est. & $95\%$ CI\\
\midrule
\showrowcolors
Intercept & 23.41 & (20.677; 26.068)\\
Time & 0.99 & (0.797; 1.178)\\
Reoperation & -12.83 & (-18.735; -6.647)\\
$t_{+}$ & -0.02 & (-0.994; 1.198)\\
Age & -0.13 & (-0.226; -0.033)\\
\addlinespace
$Gender_{Female}$ & -4.01 & (-6.671; -1.306)\\
$\sigma$ & 10.52 & (10.262; 10.8)\\
\bottomrule
\end{tabular}
\end{table}
\rowcolors{2}{white}{white}

\rowcolors{2}{gray!6}{white}
\begin{table}[!h]

\caption{\label{tab:}Estimated hazard ratios and $95\%$ credibility intervals for the parameters of the survival submodels based on the three joint models fitted to the Pulmonary Gradient dataset.}
\centering
\begin{tabular}[t]{lllllll}
\hiderowcolors
\toprule
\multicolumn{1}{c}{} & \multicolumn{2}{c}{Value} & \multicolumn{2}{c}{Value + Slope} & \multicolumn{2}{c}{Slope} \\
\cmidrule(l{2pt}r{2pt}){2-3} \cmidrule(l{2pt}r{2pt}){4-5} \cmidrule(l{2pt}r{2pt}){6-7}
  & HR & $95\%$ CI & HR & $95\%$ CI & HR & $95\%$ CI\\
\midrule
\showrowcolors
Reoperation & 0.34 & (0.056; 1.399) & 0.32 & (0.042; 1.406) & 0.31 & (0.042; 1.255)\\
$Gender_{Female}$ & 0.51 & (0.223; 1.073) & 0.52 & (0.235; 1.099) & 0.51 & (0.25; 1.061)\\
Age & 1.05 & (1.03; 1.078) & 1.06 & (1.03; 1.083) & 1.06 & (1.03; 1.08)\\
$\alpha_{1}$ & 1.01 & (0.992; 1.031) & 1.01 & (0.983; 1.032) & 1.32 & (0.783; 1.919)\\
$\alpha_{2}$ &  &  & 1.17 & (0.624; 1.987) &  & \\
\bottomrule
\end{tabular}
\end{table}
\rowcolors{2}{white}{white}

Despite the weak magnitude of the association, we can use the fitted joint models to show how individualized dynamic predictions can be derived for a new subject, under different scenarios with respect to the timing of reoperation in the future. For this illustration we will use model M3, since the slope parametrization was found to have a stronger effect on the instantaneous risk for death. In Figure 3 the red dots depict the observed longitudinal values for a new subject, the dashed vertical line shows the timing of the reoperation and the red and black solid lines denote the individual prediction of survival for this subject assuming no reoperation in the future and reoperation immediately or after four years respectively. As shown in Figure 3, reoperation improves the prediction of survival for the new subject regardless its timing. 
\linebreak 

\begin{minipage}[!h]{\linewidth}
\includegraphics[scale = 0.85]{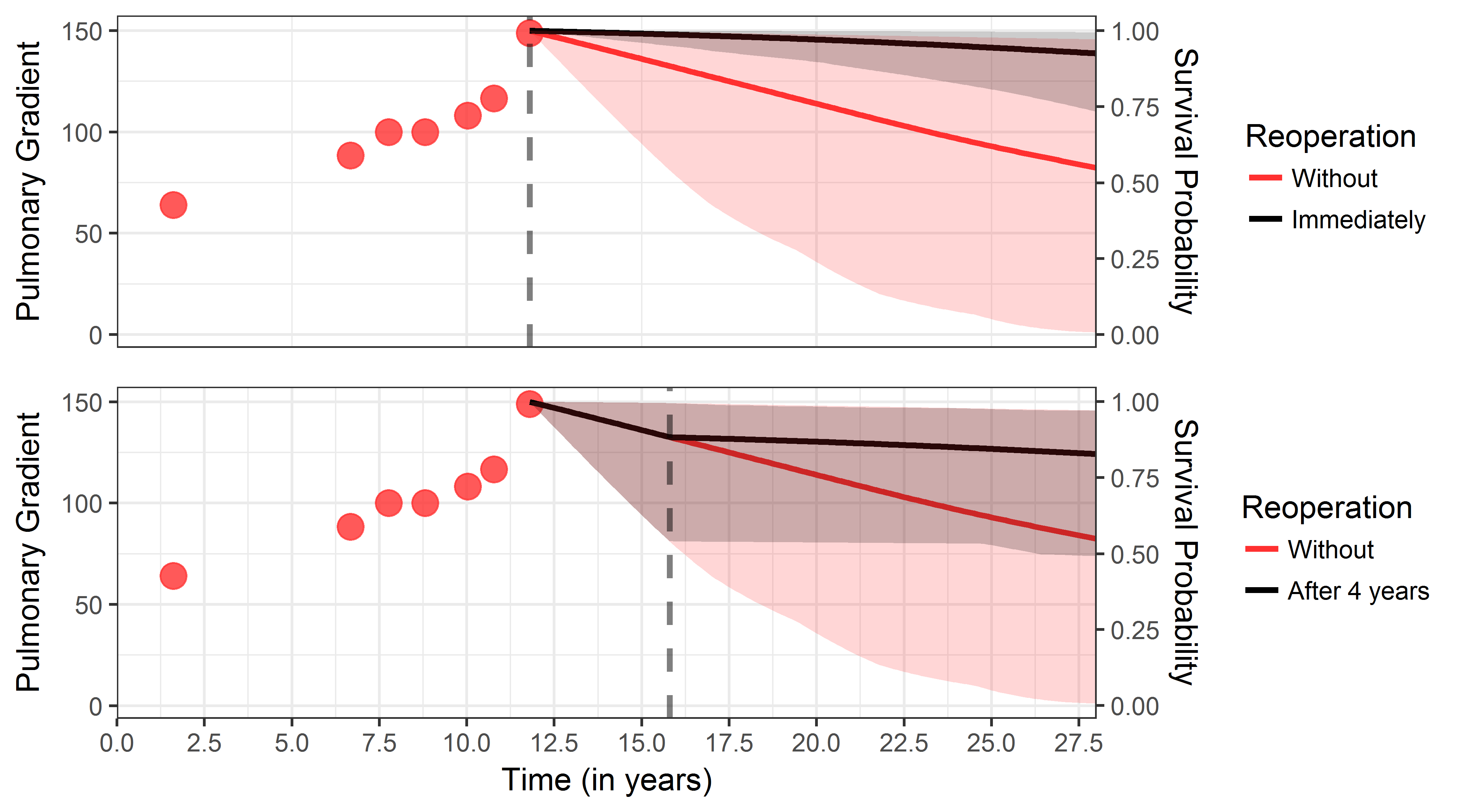}
\captionof{figure}{Survival Probabilities for a new subject under different scenarios with respect to reoperation timing.}
\end{minipage}

\subsection{SPRINT dataset}

The SPRINT dataset was also introduced in Section 1. Our goal, is to investigate how serious adverse events during follow-up change the evolution of the systolic blood pressure and the instantaneous risk for the composite endpoint, and then to utilize this information to derive individualized dynamic predictions under different scenarios with respect to the occurrence of serious adverse events in the future. 

In Figure 2, a random sample of the evolutions of systolic blood pressure for patients who experienced and who did not experience serious adverse events are depicted. For both sets of patients the profiles show diverse nonlinear trends which we assume to change after the occurrence of serious adverse events. Therefore, for this outcome, we assumed a nonlinear mixed-effects submodel using natural cubic splines with 3 knots for the effect of time, and the effect of time relative to the occurrence of serious adverse events in both the fixed-effects and random-effects parts of the model while adjusting for differences in treatment. More specifically, the following specification for the mixed-effects model was used: 

\begin{equation}\nonumber
SBP_{i}\left(t\right) = \left\{
\begin{array}{lr}
\left(\beta_{0} + b_{i0}\right) + \left(\sum_{k = 1}^{K} \left(\beta_{\left(k+1\right)} + b_{i\left(k+1\right)}\right) B_{K}\left(t, \bm{k}\right)\right) + \beta_{5}\text{Treatment} + \left(\sum_{k = 1}^{K} \beta_{\left(k+6\right)}B_{K} \left(t, \bm{k}\right)\right) \\ \times \text{Treatment} + \epsilon_{i}\left(t\right), & \hspace{0.2cm} 0 < t < \rho_{i} \vspace{0.4cm} \\
\left(\beta_{0} + b_{i0}\right) + \left(\sum_{k = 1}^{K} \left(\beta_{\left(k+1\right)} + b_{i\left(k+1\right)}\right) B_{K}\left(t, \bm{k}\right)\right) + \beta_{5}\text{Treatment} + \left(\sum_{k = 1}^{K} \beta_{\left(k+6\right)} B_{K}\left(t, \bm{k}\right)\right)\\ \times \text{Treatment} + \left(\sum_{k = 1}^{K} \left(\beta_{\left(k+10\right)} + b_{i\left(k+10\right)}\right) B_{K}\left(t_{+}, \bm{k}\right)\right) +  \left(\sum_{k = 1}^{K} \beta_{\left(k+14\right)} B_{K}\left(t_{+}, \bm{k}\right)\right)\times\\ \text{Treatment} + \epsilon_{i}\left(t\right), & \hspace{0.2cm} t \geq \rho_{i}.
\end{array}
\right.
\end{equation}

where $SBP_{i}\left(t\right)$ are the measurements of systolic blood pressure and $t_{+} = \max\left(0, t_{ij} - \rho_{ij}; j = 1, \dots, n_{i}\right)$ is the time relative to the occurrence of serious adverse event.

To investigate the association between the systolic blood pressure and the composite endpoint, we postulated relative risk submodels with different parametrizations for the systolic blood pressure. The baseline hazard was expressed as a B-splines function. We also corrected for treatment and assumed the occurrence of serious adverse event to have a direct effect on the hazard. The functional forms we used for the association structure were the current value, slope, area, both the current value and slope, as well as more elaborate ones assuming that the value, slope and area all have an effect on the hazard and assuming that after the occurrence of the serious adverse event the effect of the current slope on the hazard changes. 

Table 3 summarizes the parameter estimates and the 95\% credibility intervals of the longitudinal submodel that was used for the SPRINT dataset. Table 4 summarizes the parameter estimates and the 95\% credibility intervals of the survival submodels based on the six joint models fitted to the SPRINT dataset. As shown in Table 4, the association of the pulmonary gradient with the instantaneous risk of composite endpoint was weak in magnitude but significant in the cases of value and slope association and area association. 

\rowcolors{2}{gray!6}{white}
\begin{table}[!h]

\caption{\label{tab:}Estimated coefficients and $95\%$ credibility intervals for the parameters of the longitudinal submodel fitted to the SPRINT dataset.}
\centering
\begin{tabular}[t]{lrl}
\hiderowcolors
\toprule
  & Est. & $95\%$ CI\\
\midrule
\showrowcolors
Intercept & 137.95 & (137.549; 138.32)\\
$B_{1}\left(t, \bm{k}\right)$ & -1.33 & (-1.808; -0.851)\\
$B_{2}\left(t, \bm{k}\right)$ & -0.40 & (-0.873; 0.05)\\
$B_{3}\left(t, \bm{k}\right)$ & -8.39 & (-9.302; -7.505)\\
$B_{4}\left(t, \bm{k}\right)$ & 1.01 & (0.574; 1.407)\\
\addlinespace
$Treatment_{Intensive}$ & -0.57 & (-1.116; 0.021)\\
$B_{1}\left(t_{+}, \bm{k}\right)$ & 0.06 & (-0.954; 1.088)\\
$B_{2}\left(t_{+}, \bm{k}\right)$ & -0.06 & (-1.282; 1.181)\\
$B_{3}\left(t_{+}, \bm{k}\right)$ & -1.82 & (-3.097; -0.389)\\
$B_{4}\left(t_{+}, \bm{k}\right)$ & -0.62 & (-2.402; 1.268)\\
\addlinespace
$B_{1}\left(t, \bm{k}\right) \times Treatment_{Intensive}$ & -13.45 & (-14.125; -12.778)\\
$B_{2}\left(t, \bm{k}\right) \times Treatment_{Intensive}$ & -11.10 & (-11.734; -10.456)\\
$B_{3}\left(t, \bm{k}\right) \times Treatment_{Intensive}$ & -25.93 & (-27.184; -24.583)\\
$B_{4}\left(t, \bm{k}\right) \times Treatment_{Intensive}$ & -9.51 & (-10.094; -8.938)\\
$B_{1}\left(t_{+}, \bm{k}\right) \times Treatment_{Intensive}$ & 1.84 & (0.423; 3.333)\\
\addlinespace
$B_{2}\left(t_{+}, \bm{k}\right) \times Treatment_{Intensive}$ & 0.35 & (-1.321; 2.107)\\
$B_{3}\left(t_{+}, \bm{k}\right) \times Treatment_{Intensive}$ & 5.06 & (2.983; 6.847)\\
$B_{4}\left(t_{+}, \bm{k}\right) \times Treatment_{Intensive}$ & 1.48 & (-1.028; 3.808)\\
$\sigma$ & 11.22 & (11.164; 11.266)\\
\bottomrule
\end{tabular}
\end{table}
\rowcolors{2}{white}{white}

\rowcolors{2}{gray!6}{white}
\begin{table}[!h]

\caption{\label{tab:}Estimated hazard ratios and $95\%$ credibility intervals for the parameters of the joint models fitted to the SPRINT dataset.}
\centering
\resizebox{\linewidth}{!}{\begin{tabular}[t]{lllllllllllll}
\hiderowcolors
\toprule
\multicolumn{1}{c}{} & \multicolumn{2}{c}{Value} & \multicolumn{2}{c}{Slope} & \multicolumn{2}{c}{Value + Slope} & \multicolumn{2}{c}{Area} & \multicolumn{2}{c}{Value + Slope + Area} & \multicolumn{2}{c}{Value  + Slope Int} \\
\cmidrule(l{2pt}r{2pt}){2-3} \cmidrule(l{2pt}r{2pt}){4-5} \cmidrule(l{2pt}r{2pt}){6-7} \cmidrule(l{2pt}r{2pt}){8-9} \cmidrule(l{2pt}r{2pt}){10-11} \cmidrule(l{2pt}r{2pt}){12-13}
  & HR & $95\%$ CI & HR & $95\%$ CI & HR & $95\%$ CI & HR & $95\%$ CI & HR & $95\%$ CI & HR & $95\%$ CI\\
\midrule
\showrowcolors
$Treatment_{Intensive}$ & 0.91 & (0.59; 0.859) & 0.72 & (0.59; 0.859) & 0.9 & (0.743; 1.093) & 0.81 & (0.69; 0.97) & 0.9 & (0.768; 1.054) & 0.9 & (0.737; 1.116)\\
Serious Adverse Event & 1.86 & (1.592; 2.379) & 1.93 & (1.592; 2.379) & 1.89 & (1.536; 2.294) & 1.88 & (1.567; 2.236) & 1.87 & (1.649; 2.154) & 1.58 & (1.203; 2.065)\\
$\alpha_{1}$ & 1.01 & (0.997; 1.014) & 1.01 & (0.997; 1.014) & 1.02 & (1.008; 1.025) & 1 & (1; 1.007) & 1.01 & (1.008; 1.02) & 1.02 & (1.006; 1.026)\\
$\alpha_{2}$ &  &  &  &  & 1 & (1.001; 1.009) &  &  & 1 & (1.002; 1.006) & 1 & (0.998; 1.007)\\
$\alpha_{3}$ &  &  &  &  &  &  &  &  & 1 & (0.999; 1.003) & 1.03 & (0.999; 1.048)\\
\bottomrule
\end{tabular}}
\end{table}
\rowcolors{2}{white}{white}

Similarly, as for the Pulmonary Gradient Dataset, in Figure 4 we illustrate how the individualized subject-specific predictions for the composite endpoint of interest change under different scenarios for the timing of the occurrence of a serious adverse event. More specifically, we illustrate the cases of an immediate occurrence of the adverse event and an occurrence after a year. In both cases the occurrence of the serious adverse event worsens the survival prediction. 
\vspace{0.4cm} 

\begin{minipage}[!h]{\linewidth}
\includegraphics[scale = 0.85]{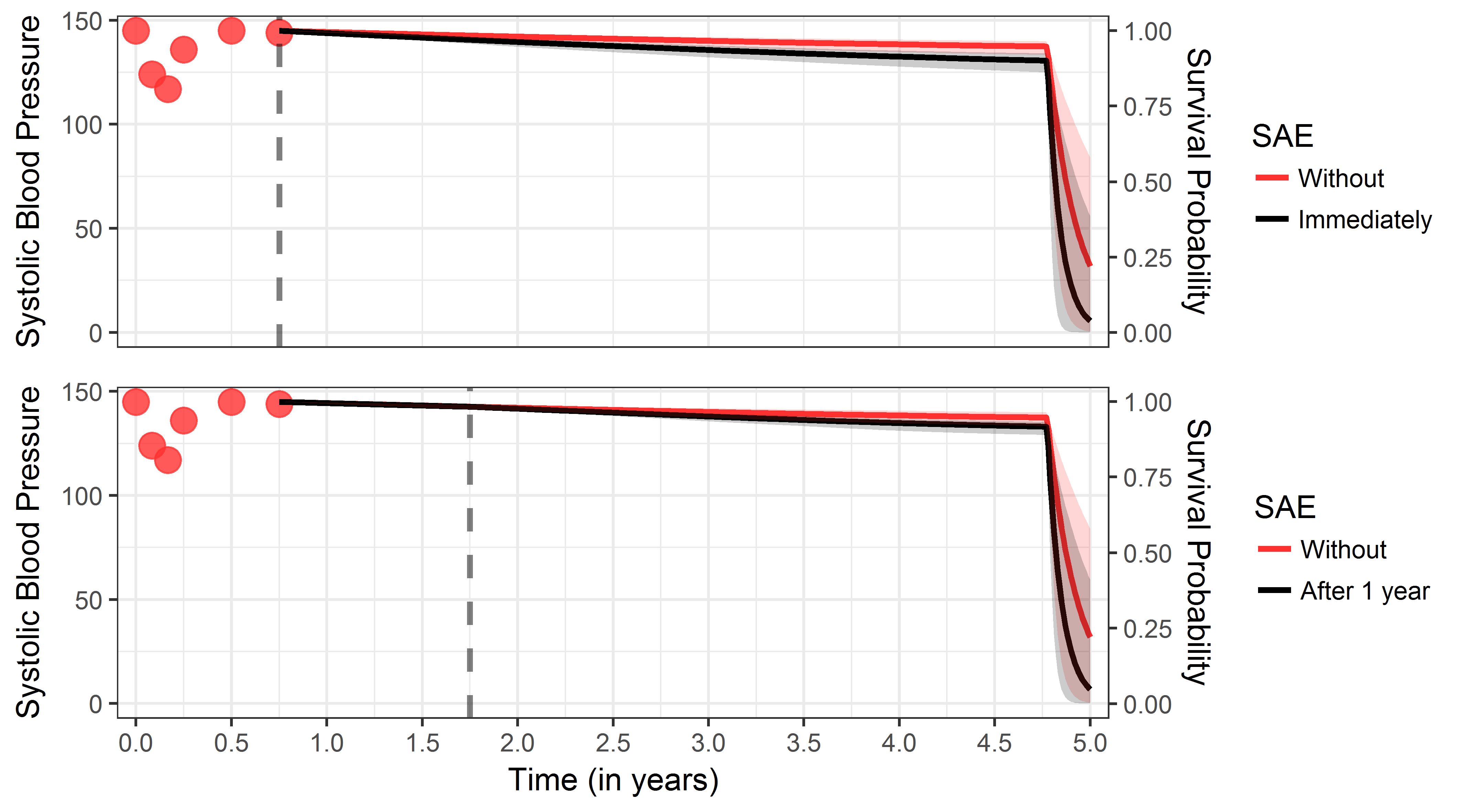}
\captionof{figure}{Survival Probabilities for a new subject under different scenarios with respect to the occurence of serious adverse event (SAE).}
\end{minipage}

\section{Simulation Study}

\subsection{Design}

To evaluate the performance of the proposed models and to compare, in terms of predictive accuracy, the dynamic predictions that account for the whole biomarker trajectory against the case where extrapolation is assumed, we performed a simulation study. The main goal of the simulation study is to show the benefit in the accuracy of the individualized dynamic predictions when assuming that the intermediate event changes both the risk for the event of interest and the longitudinal trajectory against the case assuming that the intermediate event only changes the risk for the event of interest while the longitudinal trajectory is extrapolated. We assumed $1500$ patients and planned follow-up visits randomly selected from $0$ to $50$. To mimic a realistic situation, the timing of the intermediate event was assumed to depend on the value of the biomarker trajectory. Specifically, if the biomarker exceeded a specific value then reintervention took place at the next visit. For the cases that this value was not reached, the patient was assumed to never have experienced the intermediate event. For simplicity we assumed a linear mixed-effects model and a survival submodel without any baseline covariates.

For the continuous longitudinal outcome, we simulated the data from a linear mixed-effects models similar to the model that we used for the pulmonary gradient dataset:

\begin{equation}\nonumber
y_{i}\left(t\right) = \eta_{i}\left(t\right) + \epsilon_{i}\left(t\right) = \left(\beta_{0} + b_{i0}\right) + \left(\beta_{1} + b_{i1}\right)t + \left(\tilde{\beta}_{3} + \tilde{b}_{i3}\right)R_{i}\left(t\right) + \left(\tilde{\beta}_{4} + \tilde{b}_{i4}\right)t_{+} + \epsilon_{i}\left(t\right),
\end{equation}
\linebreak
where $\epsilon_{i}\left(t\right) \sim \mathcal{N}\left(0, \sigma^{2}I_{n_{i}}\right)$ and $\bm{b} \sim \mathcal{N}\left(0, \bm{D}\right)$. More specifically, we adopted a linear effect for time, a "drop" effect that occurs at the time of reoperation and an effect for the change in the slope from the time of reoperation onwards for both the fixed and the random part. Time $t$ was simulated from a uniform distribution between $0$ and $50$. Based on this model for the continuous outcome we assumed three different scenarios: 

\begin{itemize}
\item{$\text{Scenario 1:}\hspace{0.2cm} \beta_{1} = 20.7, \hspace{0.2cm} \beta_{2} = 1.6,\hspace{0.2cm} \tilde{\beta}_{3} = -15.5,\hspace{0.2cm} \tilde{\beta}_{4} = -0.76,$}
\item{$\text{Scenario 2:}\hspace{0.2cm} \beta_{1} = 20.7,\hspace{0.2cm} \beta_{2} = 1.6,\hspace{0.2cm} \tilde{\beta}_{3} = -15.5,\hspace{0.2cm} \tilde{\beta}_{4} = 0,$}
\item{$\text{Scenario 3:}\hspace{0.2cm} \beta_{1} = 20.7,\hspace{0.2cm} \beta_{2} = 1.6,\hspace{0.2cm} \tilde{\beta}_{3} = 0,\hspace{0.2cm} \tilde{\beta}_{4} = -0.76.$}
\end{itemize}
\noindent
The assumed longitudinal trajectories for each of the three scenarios are depicted in Figure 3.
\linebreak 

\begin{minipage}{\linewidth}
\includegraphics{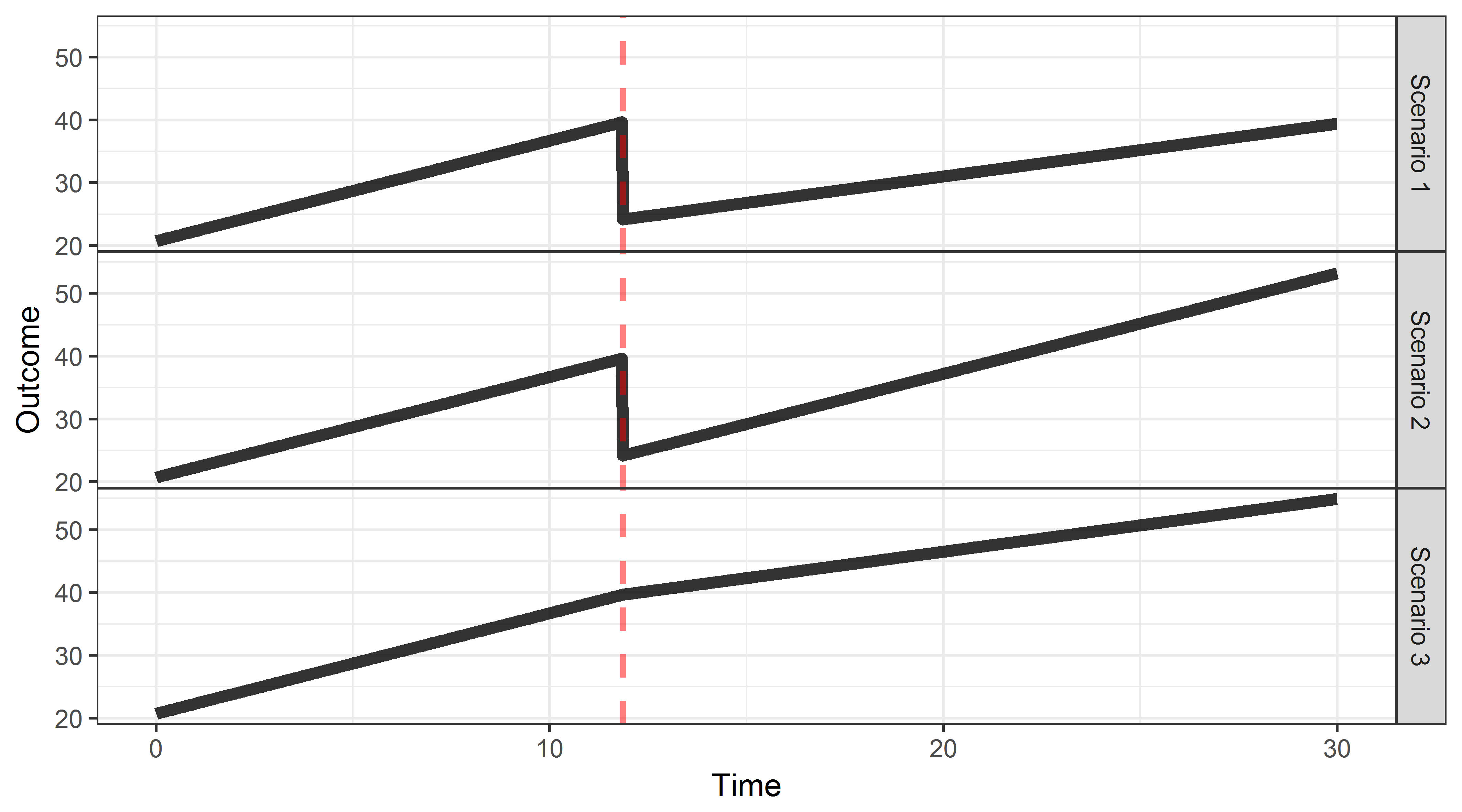}
\captionof{figure}{Assumed average longitudinal evolutions under the three simulation scenarios.}
\end{minipage}
\linebreak

More specifically, in the first scenario we assume that the longitudinal profile drops at the occurrence of the intermediate while the slope changes after its occurrence. In the second and third scenarios the slope does not change after the occurrence of the intermediate event and the longitudinal profile does not drop respectively. 

For the survival outcome we assumed a relative risk model of the form:

\begin{equation}\label{eq:11}
h_{i}\left(t\right) = h_{0}\left(t\right)\exp\{R_{i}\left(t\right)\zeta + \alpha_{1}\eta_{i}\left(t\right)\},
\end{equation}
\linebreak
where the baseline risk was simulated from a Weibull distribution $h_{0}\left(t\right) = \xi t^{\xi-1}$ with $\xi = 20.4$. The censoring process was assumed to follow an exponential distribution with mean equal to $22.6$.

\subsection{Results}

Under the settings described in the previous Section, 250 datasets were simulated for each of the three scenarios. All the datasets were split in half to a training and test part with $750$ subjects each. For all the scenarios, to account for the whole trajectory of the biomarker, the joint model which consists of the submodels in \eqref{eq:1} and \eqref{eq:2} was fitted to the part of the simulated datasets which were kept for training. On the other hand, for the extrapolation method the observations after reintervention were omitted from the analysis of the longitudinal outcome and the following mixed-effects model was fitted to the data up to reintervention time: 

\begin{equation}\label{eq:12}
y_{i}\left(t\right) = \eta_{i}\left(t\right) + \epsilon_{i}\left(t\right) = \left(\beta_{0} + b_{i0}\right) + \left(\beta_{1} + b_{i1}\right)t + \epsilon_{i}\left(t\right),
\end{equation}
\linebreak
while for the survival process the same model was used.

To assess the performance of the two approaches we used the models that were fitted on the training data to calculate the time-dependent AUCs and the prediction errors based on the test data. Both the time-dependent AUCs and the prediction errors were calculated at $3$ different time-intervals starting at: $t = 20$, $t = 22$ and $t = 24$ respectively and assuming a clinically relevant time interval of two years $\Delta t = 2$. The time-intervals were selected on the basis of when the most events occur in the simulated data.   

In Figures 4 and 5 we present the results of the simulation study depicted by boxplots. Specifically, the boxplots in each row represent different scenarios: non-zero effects, zero change in slope and zero "drop" at time of reintervention and in each column a different time-interval for prediction. In all the scenarios and time-intervals, both the AUC and PE are better when assuming that the intermediate event changes both the risk for the event of interest and the longitudinal trajectory. Moreover, there is a slight increase in the difference between the two methods for both predictive measures as the time-interval is set at later time-points. That is, the more information used the greater the difference between the two methods tends to become. Moreover, the relative performance of the two approaches does not differ between the three scenarios as well as for the different follow-up times. Therefore, the results support the argument that accounting for the changes in the longitudinal trajectories due to the occurrence of the intermediate event improve the predictive accuracy when compared to the approach that the longitudinal profile remains unaffected by its occurrence. 
\linebreak

\begin{minipage}{\linewidth}
\includegraphics{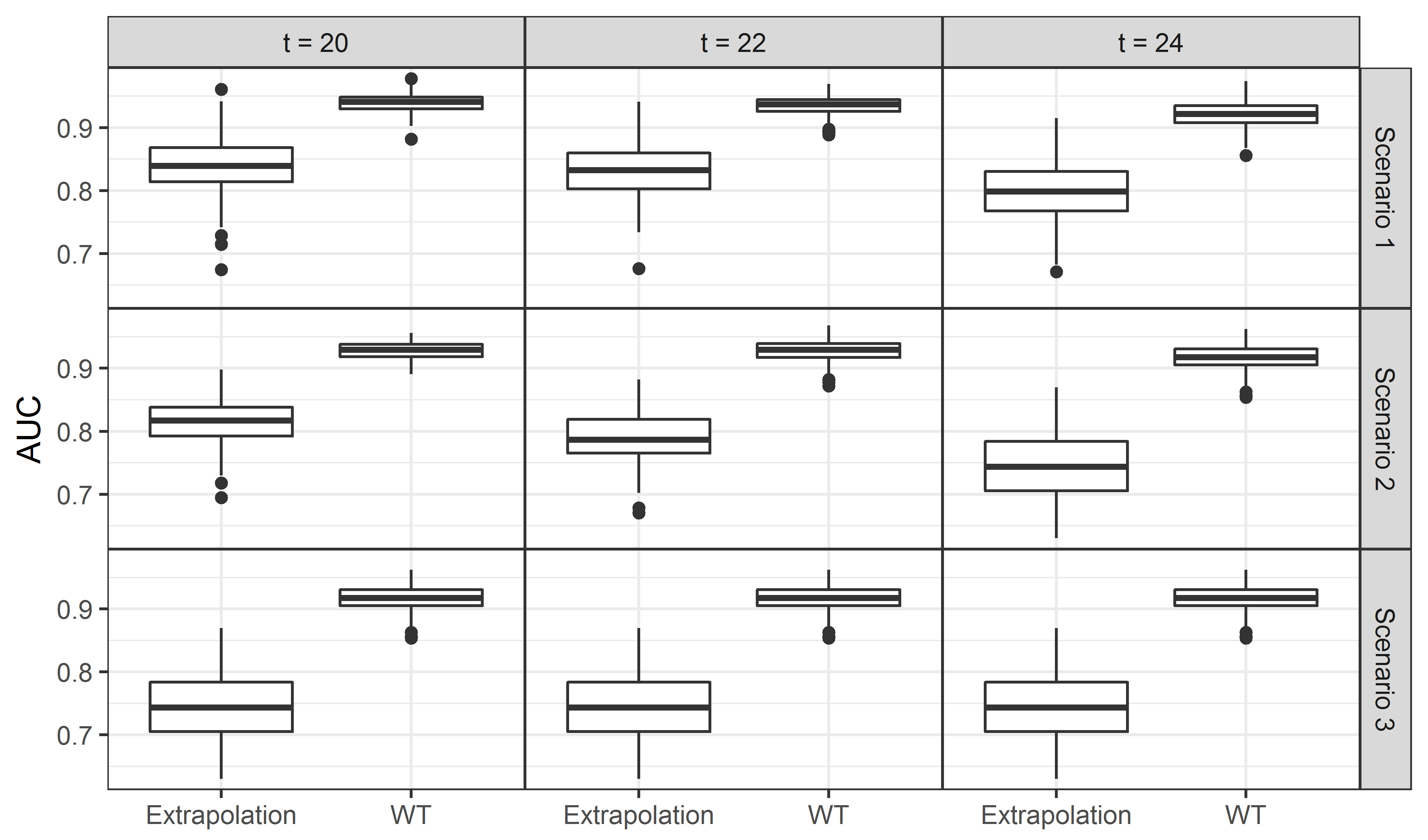}
\captionof{figure}{AUCs for the individualized dynamic predictions, evaluated using the testing part of the 250 datasets for two different joint models. $\left.1\right)$ Extrapolation: Assuming that the longitudinal profile does not change after the occurrence of the intermediate event, $\left.2\right)$ WT: Assuming that the longitudinal profile changes after the occurrence of the intermediate event.}
\end{minipage}

\begin{minipage}{\linewidth}
\includegraphics{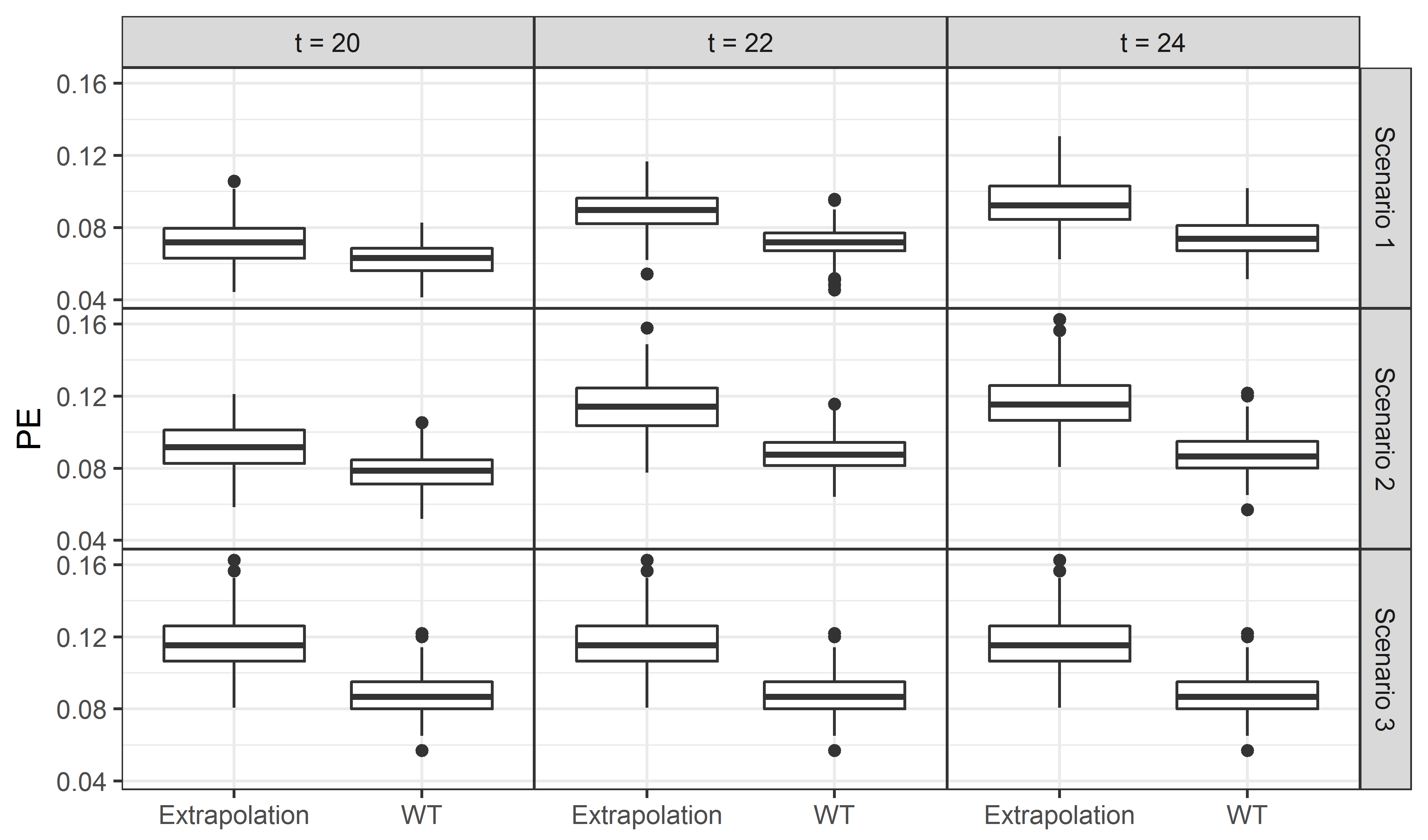}
\captionof{figure}{PEs for the individualized dynamic predictions, evaluated using the testing part of the 250 datasets for two different joint models. $\left.1\right)$ Extrapolation: Assuming that the longitudinal profile does not change after the occurrence of the intermediate event, $\left.2\right)$ WT: Assuming that the longitudinal profile changes after the occurrence of the intermediate event.}
\end{minipage}

\section{Discussion}

Using the joint modeling framework we developed tools for deriving individualized dynamic predictions that are adaptive to different scenarios regarding intermediate events, such as treatment changes or the occurrence of adverse events. We proposed a range of joint models for longitudinal and time-to-event data which can accommodate special features due to the occurrence of intermediate events in both the longitudinal and survival submodels. That is, by incorporating features, such as the ones described in \eqref{eq:1} and \eqref{eq:2} a broad range of flexible joint models is sketched which accounts for the impact of an intermediate event by allowing for: $\left.1\right)$ a direct effect of the intermediate event on the risk of the clinical endpoint through the time-varying binary covariate $R_{i}\left(t\right)$, $\left.2\right)$ a direct effect of the intermediate event on the longitudinal trajectory through $g\{R_{i}\left(t\right), t_{i+}\}$ and $\left.3\right)$ an indirect effect of the intermediate event on the the risk of the clinical endpoint through the association between the two outcomes which is defined by $f_{\left(t, \rho_{i}\right)}\{(\mathcal{H}_{i}\left(t, \rho_{i}\right), \bm{b}_{i}\}$ and is allowed to differ before and after the intermediate event. All these features allow for great flexibility in the specification of the joint model which when utilized accordingly can lead to accurate predictions. 

In the same line as recent observations with regard to dynamic predictions from joint models, we have seen that the accuracy of the predictions is influenced by intermediate events occurring during follow-up. Such events will need to be appropriately modeled as time-varying covariates in both the longitudinal and survival submodels. As illustrated in our simulation study, doing so, improves the predictive accuracy of the individualized dynamic predictions. 

The joint model formulation we presented allows to utilize the quantification of the effects the intermediate event imposes on the risk for the clinical endpoint of interest. As such, it can be utilized to derive individualized dynamic predictions for new subjects who did not experience the intermediate event and quantify how its occurrence at any future time point will influence their risk predictions. Such a predictive tool can provide valuable information to the physicians and assist in their decision making process for potential treatment changes. Based on such predictions further prognostic tools can potentially be developed. For example in settings where the timing of a future treatment is important, having both benefits and disadvantages when applied either too late or too early, such dynamic predictions that are adaptive to the timing of the intermediate can become the basis for methodology that can be used to predict the optimal time for the future treatment. Unfortunately, the applications at hand did not allow for exploring such a possibility, but this is a clear direction for future research. Moreover, in our paper we only consider the joint analysis of one longitudinal and one survival outcome. While the extension of the proposed models to their multivariate counterparts is straightforward, such multivariate joint models have not been explored in the literature in the context of intermediate events that may occur during follow-up and alter the course of the disease for the patient. Therefore, individualized dynamic predictions based on even more complex joint models, such as with multiple longitudinal biomarkers or with multi-state processes instead of a single time-to-event outcome might lead to improved accuracy depending on the application.

\renewcommand\refname{References}

\bibliographystyle{abbrvnat} % or try abbrvnat or unsrtnat or plainnat
\bibliography{References} % refers to bibliography_arsia.bib

\end{document}